\documentclass{PoS}

\title{ANTARES search for high-energy neutrinos\\ from TeV-emitting blazars, Markarian 421 and 501,\\ in coincidence with HAWC gamma-ray flares}

\ShortTitle{ANTARES search for neutrinos from blazars, Mrk 421 and Mrk 501.}

\author{The ANTARES$^\ddagger$ and HAWC$^\dagger$ Collaborations\thanks{For collaboration list, see PoS(ICRC2019) 1177.}\\
$^\ddagger$\href{http://antares.in2p3.fr/Collaboration/index2.html}{http://antares.in2p3.fr/Collaboration/index2.html}\\
$^\dagger$\href{https://www.hawc-observatory.org/collaboration/icrc2019.php}{https://www.hawc-observatory.org/collaboration/icrc2019.php}\\
E-mail: \email{mukharbek.organokov@iphc.cnrs.fr}, \\ \hspace{31pt} \email{thierry.pradier@iphc.cnrs.fr}
}

\abstract{
An updated analysis of a targeted search for high-energy neutrinos from Markarian 421 and Markarian 501 is reported. They are two of the closest and brightest extragalactic sources in the TeV band. In contrast to other types of active galactic nuclei, BL Lacs are characterized by rapid and large-amplitude flux variability. Such radio-loud active galactic nuclei are candidate sources of the observed high-energy cosmic rays. Because their jet is collimated to our line of sight, the hadronic interactions with the surrounding medium can produce an accompanying neutrino and gamma-ray flux. The recent detection of high-energy neutrinos from the direction of TXS~0506+056 motivates a search for high-energy neutrinos from blazars with enhanced gamma-ray activity. These two targeted blazars are subject to long-term monitoring campaigns by the HAWC TeV gamma-ray observatory located in Mexico. This contribution presents the latest results of a search and extends previously presented results to a longer period that covers ANTARES data collected between November 2014 and December 2017. The gamma-ray light curves of each source were used to search for temporally correlated neutrinos, potentially produced in hadronic processes.\\

\textbf{Corresponding authors:} \speaker{Mukharbek Organokov}$^{1}$ and Thierry Pradier$^{1}$\\
$^{1}$Universit\'e de Strasbourg, CNRS, IPHC UMR 7178, F-67000 Strasbourg, France\\
}

\FullConference{36th International Cosmic Ray Conference -ICRC2019-\\
		July 24th - August 1st, 2019\\
		Madison, WI, U.S.A.}

\usepackage{tikz}

\begin{document}

\section{Introduction}
\label{sec:intro}
In this proceedings, we present the results of an  ANTARES~\cite{Collaboration:2011nsa,ICRC2019HighlightTalkConiglione} search for high-energy neutrino ($\nu$) counterparts to $\gamma$-ray emission from blazars, Mrk 421 and Mrk 501, the only two in the 2HWC catalog~\cite{Abeysekara:2017hyn} that have confirmed extragalactic associations. This study focuses on the search for space/time correlation between $\nu$-s detected by ANTARES and $\gamma$-ray flares detected by HAWC~\cite{Abeysekara:2017hyn,Albert:2017uzh} from these blazars. The very high-energy (VHE; 0.1-100 TeV) extragalactic sky is dominated by emission from blazars~\cite{Cerruti:2016msf}, a class of a radio-loud AGN~\cite{Beckmann:2013wte}. Two blazars, Mrk 421 and Mrk 501, are the brightest and closest BL Lac objects known, at luminosity distances $\mathrm{d_L}$ = 134 Mpc with redshift z = 0.031 and $\mathrm{d_L}$ = 143 Mpc with redshift z=0.033 respectively. These blazars are the first~\cite{Punch:1992xw} and the second~\cite{Quinn:1996dj} extragalactic objects discovered in the TeV energy band; thus, are of high good candidates for a $\nu$ counterpart. The latest results of a search which extends previously presented results~\cite{Organokov:2018tmu} to a longer period that covers ANTARES data collected between November 2014 and December 2017 are discussed. As the nearest blazars to Earth, both are excellent sources to test the blazar-neutrino connection scenario, especially during flares where time-dependent $\nu$ searches may have a higher detection probability because of a reduced background~\cite{Organokov:2018tmu,Pradier:2017kcp,Albert:2017uld}.

\section{ANTARES and HAWC}
\label{sec:antaresandhawc}
The ANTARES (Astronomy with a Neutrino Telescope and Abyss environmental RESearch) $\nu$ telescope~\cite{Collaboration:2011nsa} is a \v{C}herenkov $\nu$ detector in the Mediterranean Sea and most sensitive for $\nu$ energies 100 GeV $<E_{\nu}<$ 100 TeV. The HAWC (High-Altitude Water Cherenkov Observatory)\footnote{HAWC Collaboration, \href{https://www.hawc-observatory.org/}{https://www.hawc-observatory.org/}} $\gamma$-ray observatory~\cite{Albert:2017uzh} is a \v{C}herenkov detector designed to search for VHE $\gamma$-rays  (100 GeV $<E_{\gamma}<$ 100 TeV) from astrophysical sources by detecting the \v{C}herenkov light emission from charged particles in $\gamma$-ray induced air showers. HAWC is located at an elevation of 4,100 m above sea level on the flanks of the Sierra Negra volcano in the state of Puebla, Mexico ($18^\circ 99'$ N, $97^\circ 18'$ W). HAWC is the most sensitive wide field-of-view TeV telescope currently in operation~\cite{Abeysekara:2017hyn}.

\subsection{The ANTARES data set}
\label{sec:DATAant}
The ANTARES data set covers the same period of observation as HAWC: 
\begin{itemize}
    \item Mrk 421: November $27^{\mathrm{th}}$, 2014 $-$ December $31^{\mathrm{st}}$, 2018 (MJD: 56988-58119)
    \item Mrk 501: November $28^{\mathrm{th}}$, 2014 $-$ December $26^{\mathrm{th}}$, 2016 (MJD: 56989-57567)
\end{itemize}
This leads to an effective detector livetimes: 1099.93/561.55 days for Mrk 421/501 respectively.

The search relies on track-like event signatures, so only charged-current (CC) interactions of muon neutrinos are considered. The muon ($\mu$) track reconstruction returns two quality parameters, namely the track-fit quality parameter, $\Lambda$, and the estimated angular uncertainty on the fitted $\mu$ track direction, $\beta$~\cite{Organokov:2018tmu}. Cuts on these parameters are used to improve the signal-to-noise ratio~\cite{Organokov:2018tmu}. Atmospheric $\nu$-s are the main source of background; an additional source of background is due to the mis-reconstructed atmospheric $\mu$-s~\cite{Pradier:2017kcp,Albert:2017uld}. 

\subsection{The HAWC light curves of the two blazars}
\label{sec:LChawc}
\begin{figure}[ht!]
\vspace{-25pt}
\centering

  \begin{tikzpicture}
    \node[anchor=south west,inner sep=0] (image) at (0,0) {\includegraphics[width=1.00\textwidth]{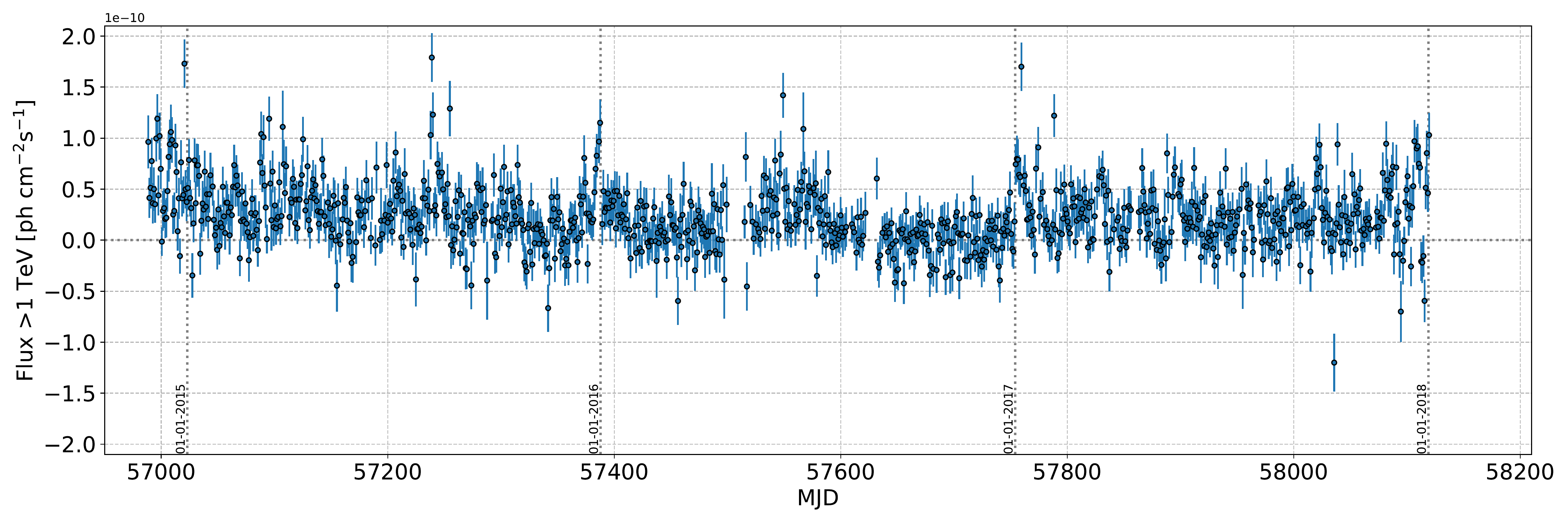} };
    \begin{scope}[x={(image.south east)},y={(image.north west)}]

        \node[text width=8cm, minimum height=3cm,minimum width=8cm] at (0.945,0.88) { \textbf{ \textcolor{red}{
        \normalsize{PRELIMINARY}  
       } } };
       \node[text width=8cm, minimum height=3cm,minimum width=8cm] at (0.39,0.2) { \textbf{ \textcolor{blue}{
        \large{Mrk 421}  
       } } };
       \end{scope}
   \end{tikzpicture}\\
   \vspace{-25pt}
     \begin{tikzpicture}
    \node[anchor=south west,inner sep=0] (image) at (0,0) {\includegraphics[width=1.00\textwidth]{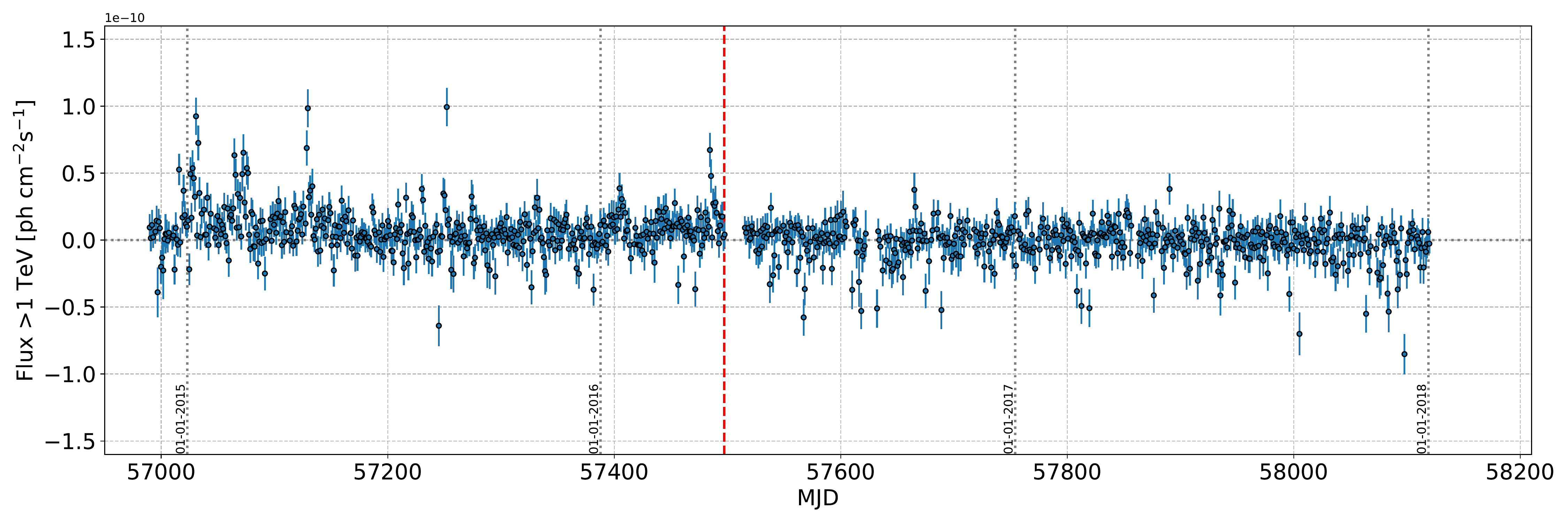} };
    \begin{scope}[x={(image.south east)},y={(image.north west)}]

        \node[text width=8cm, minimum height=3cm,minimum width=8cm] at (0.945,0.86) { \textbf{ \textcolor{red}{
        \normalsize{PRELIMINARY}  
       } } };
       \node[text width=8cm, minimum height=3cm,minimum width=8cm] at (0.39,0.22) { \textbf{ \textcolor{blue}{
       \large{Mrk 501}  
       } } };
       \end{scope}
   \end{tikzpicture}
    \caption{The LCs of both blazars credited by HAWC. The red dashed line represent the date outside of which there is no activity found (flat long block has been identified), hence the data is not used.}
    \label{fig:LCs}
\end{figure}
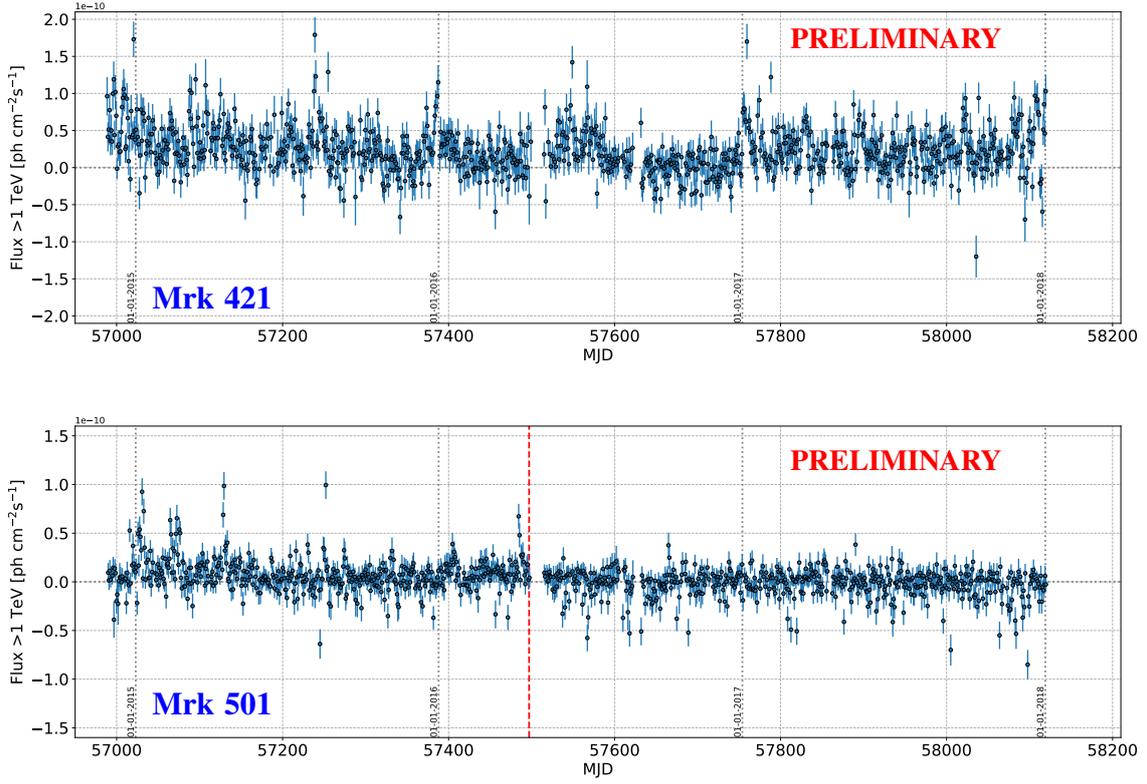

Taking advantage of $\gamma$-ray flux time variation information from potential $\nu$ emitters, the $\nu$ background can be significantly reduced and signal-to-noise discrimination improved. HAWC has made clear detections of Mrk 421 and Mrk 501 and the last HAWC 1017 day light curves (LCs) of these blazars (see Fig.~\ref{fig:LCs}) are used to determine the periods of interest for the coincident $\nu$ search. The signal time probability density functions (PDF) is assumed to have a square shape. The precise shape of the signal time PDF can be extracted directly for the $\gamma$-ray LC assuming the proportionality between the $\gamma$-ray and the $\nu$ fluxes. 

The Bayesian blocks algorithm~\cite{Scargle:2012gq} can be applied to detect and characterize signals in noisy time series such as LCs. If an LC is variable, the Bayesian blocks algorithm can be used to find optimal data segmentation into regions that are well represented by constant flux, within the statistical uncertainties~\cite{Albert:2017uzh}. It identifies the changes between flux states via finding points at the transition from one flux state to another one and as a result distinguishes distinct flux states~\cite{Albert:2017uzh}. Both Mrk 421 and Mrk 501 show clear variability on time scales of one day~\cite{Albert:2017uzh}. The 1-day binning is applied to the final distinct fluxes from and used in the analysis as a signal time PDF.

In addition, several flare selection conditions are considered: all flare states are taken as they are (long case); only those flare states are taken which pass the defined \textit{average flux}+$\mathit{2\sigma}$ threshold (short case). Earlier investigations~\cite{Organokov:2018tmu} show that \textit{average flux}+$\mathit{2\sigma}$ threshold is sufficient from a variety of thresholds have been tested such as \textit{average flux}, \textit{average flux}+$\mathit{1\sigma}$, \textit{average flux}+$\mathit{2\sigma}$ due to the shorter lengths of the flaring period that rejects more efficiently the background.

\section{Search Method}
\label{sec:method}
A search for $\nu$ candidades in coincidence with $\gamma$-rays from astrophysical sources is performed using an unbinned likelihood-ratio maximization method~\cite{Adrian-Martinez:2015wis,Albert:2016gtl,Pradier:2017kcp,Albert:2017uld}. The goal is to determine the relative contribution of background and signal components for a given direction in the sky and at a given time.
\begin{equation}
\centering
\mathrm{ln(\mathsf{L})}=\bigg(\sum_{\mathrm{i=1}}^{\mathrm{N}} \mathrm{ln[N_S S_i + N_B B_i]}\bigg) - \mathrm{[N_S + N_B]}
\end{equation} 

To perform the analysis, the ANTARES data sample is parametrized as two-component mixture of signal and background. Since the signal is expected to be small, the total number of events N in the considered data sample can be treated as background. $\mathrm{S_i}$ and $\mathrm{B_i}$ are defined as the PDFs respectively for signal and background for an event i, at time $\mathrm{t_i}$, energy $\mathrm{E_i}$, declination $\delta_\mathrm{i}$~\cite{Organokov:2018tmu}. As a result, $\mathrm{S_i}$ = $\mathrm{P_s}(\alpha_i) \cdot \mathrm{P_s(E_i)} \cdot \mathrm{P_s(t_i)}$ and $\mathrm{B_i}$ = $\mathrm{P_b(sin(\delta_i))} \cdot \mathrm{P_b(E_i)} \cdot \mathrm{P_b(t_i)}$~\cite{Organokov:2018tmu}. The parameter $\alpha_i$ represents the angular distance between the direction of the event i and direction to the source~\cite{Organokov:2018tmu}. Additionally, $\mathrm{N_{S}}$ and $\mathrm{N_{B}}$ are unknown signal events and known background rate (a priori when building $\mathsf{L}$) respectively~\cite{Organokov:2018tmu}. The $\mathrm{N_{S}}$ is fitted by maximizing the likelihood $\mathsf{L}$. The background PDFs are all computed using data only~\cite{Pradier:2017kcp,Albert:2017uld}. 

\subsection{Ingredients}
\label{sec:ing}

One of the most important parameter in point source search is the angular distance of the $\nu$ events to the source, characterized by the Point Spread Function (PSF), which is defined as the probability density to find a reconstructed event at an angular distance $\Delta \Psi$ around the direction of the source~\cite{Albert:2017ohr}. The PSF - $\mathrm{P_s}(\alpha_\mathrm{i})$, can be expressed as the probability density of $\alpha$ per unit solid angle, $\Omega$~\cite{Pradier:2017kcp,Albert:2017uld}:
\begin{equation}
\centering
\mathrm{P_s}(\alpha_\mathrm{i})=\frac{\mathrm{dP}}{\mathrm{d}\Omega}=\frac{\mathrm{d}\alpha_\mathrm{i}}{\mathrm{d}\Omega}\frac{\mathrm{dP}}{\mathrm{d}\alpha_\mathrm{i}}=\frac{1}{2\pi \mathrm{sin}\alpha_\mathrm{i}} \frac{\mathrm{dP}}{\mathrm{d}\alpha_\mathrm{i}},
\label{eq:eq_psf}
\end{equation}
where $\alpha_i = |\alpha^{true}_i - \alpha^{rec}_i |$ is the difference between simulated $\nu$ direction $\alpha^{true}_i$ and reconstructed $\mu$ direction $\alpha^{rec}_i$. 

The energy PDF for the signal events is produced according to the studied energy spectra~\cite{Organokov:2018tmu}: E$^{-2.0}$, E$^{-2.5}$, E$^{-1.0} \, \mathrm{exp}\left(-\mathrm{E} \slash 1 \, \mathrm{PeV}\right)$ for both sources and extra E$^{-2.25}$ for Mrk 501 (as from the fit of the spectral shape of this source performed for same data in~\cite{deLeon:2017tbd}). It is worth noting that Mrk 421 is well described by the E$^{-2.0}$ spectrum~\cite{Organokov:2018tmu}. 

As stated above, the signal time PDF shape is extracted directly from the $\gamma$-ray LC assuming a proportionality between the $\gamma$-ray and the $\nu$ fluxes.

\subsection{Optimization}
\label{sec:optim}
To avoid biasing the analysis, it has been performed according to strict ''blinding'' policy defined by the ANTARES Collaboration. The data have been blinded by scrambling of the event right ascension (RA): the true RA of an event is hidden during the optimization steps of the analysis. This procedure avoids the selection procedure becoming inadvertently tuned toward a discovery. The ''blinding'' policy requires the simulation of a large number of pseudo-experiments (PEX), with the generation of both signal and background events. In this study, $3\times10^{5}$ PEXs are generated for background and $3\times10^{4}$ PEXs including a signal as well (from 1 up to 20 signal events are injected) in a $30^{\circ}$ cone around the considered source. The test statistics (TS) is evaluated as follows:
\begin{equation}
\mathrm{TS} = 2(\mathrm{ln}(\mathsf{L}^{max}_{s+b})-\mathrm{ln}(\mathsf{L}_{b}))
\end{equation}

Cuts on the cosine of the zenith angle of the reconstructed events $\mathrm{cos(\theta)>-0.1}$ and on the direction of the reconstructed events $\beta<1.0^{\circ}$ are used to improve the signal-to-noise ratio~\cite{Organokov:2018tmu}. The $\mathrm{\Lambda_{cut}}$ is optimized for each source on the basis of maximizing the Model Discovery Potential (MDP), a probability to make a discovery assuming that the model is correct~\cite{Adrian-Martinez:2013dsk}, for $3\sigma$/$5\sigma$ levels for each $\nu$ spectrum. Once the parameters are optimized, the results are obtained by looking at real data (''unblinding'' procedure). In case of no discovery made, the final upper limits (ULs) are derived from the unblinded dataset. Conversion from $\mathrm{N_{S}}$ to $F$, the equivalent source flux, is done through the acceptance of the detector~\cite{Organokov:2018tmu}.

\subsection{Systematics}
\label{sec:syst}
The possible systematics intrinsically inherent to the detector are considered. The systematics on the absolute pointing accuracy, angular resolution, and the energy resolution are applied as a correction over the simulated parameter that is obtained from a Gaussian distribution with that uncertainty as a standard deviation. Since the events are simulated in equatorial coordinates ($\delta$,RA), the systematic uncertainty in local coordinates ($\theta$,$\phi$) is considered in the PEX by determining an elevation $\theta$ and azimuth $\phi$ for that source in the moment of the day at which the event was simulated. In addition, the uncertainty on detector acceptance is taken into account.

\section{Results}
\label{sec:results}
No significant excess is found in this search. Thus, the ULs on the $\nu$ energy flux, $F$, and fluence, $\mathcal{F}$, at 90\% Confidence Level (CL) according to Neyman's method~\cite{Neyman:1937uhy} are established using 5-95\% energy bounds, $E_{min}$ and $E_{max}$, defined to contain 90\% of signal $\nu$ events:

\begin{equation}
\centering
F^{90\%CL} = \int E \, \Phi_{E} \, dE = \int E \, \Phi_{0} \, S(E) \, dE = \Phi^{90\%CL}_{0} \int E \, S(E) \, dE
\label{Eq:ULflux}
\end{equation}

\begin{equation}
\centering
\mathcal{F}^{90\%CL} = \int Fdt = F \Delta T = \Delta T \cdot \Phi^{90\%CL}_{0} \int_{E_{min}}^{E_{max}} E \, S(E) \, dE 
\label{Eq:ULfluence}
\end{equation}

Here $\Delta T$ is the livetime of the search [s]; $\Phi^{90\%CL}_{0}$ is the UL on the $\nu$ flux normalization [GeV$^{-1}$~cm$^{-2}$~s$^{-1}$].

\begin{figure}[!htb]
    \centering
    \vspace{-30pt}
    \begin{minipage}{.488\textwidth}
 
        \centering
    
        \begin{tikzpicture}
        \node[anchor=south west,inner sep=0] (image) at (0,0) 
        {\includegraphics[width=1.0\textwidth]{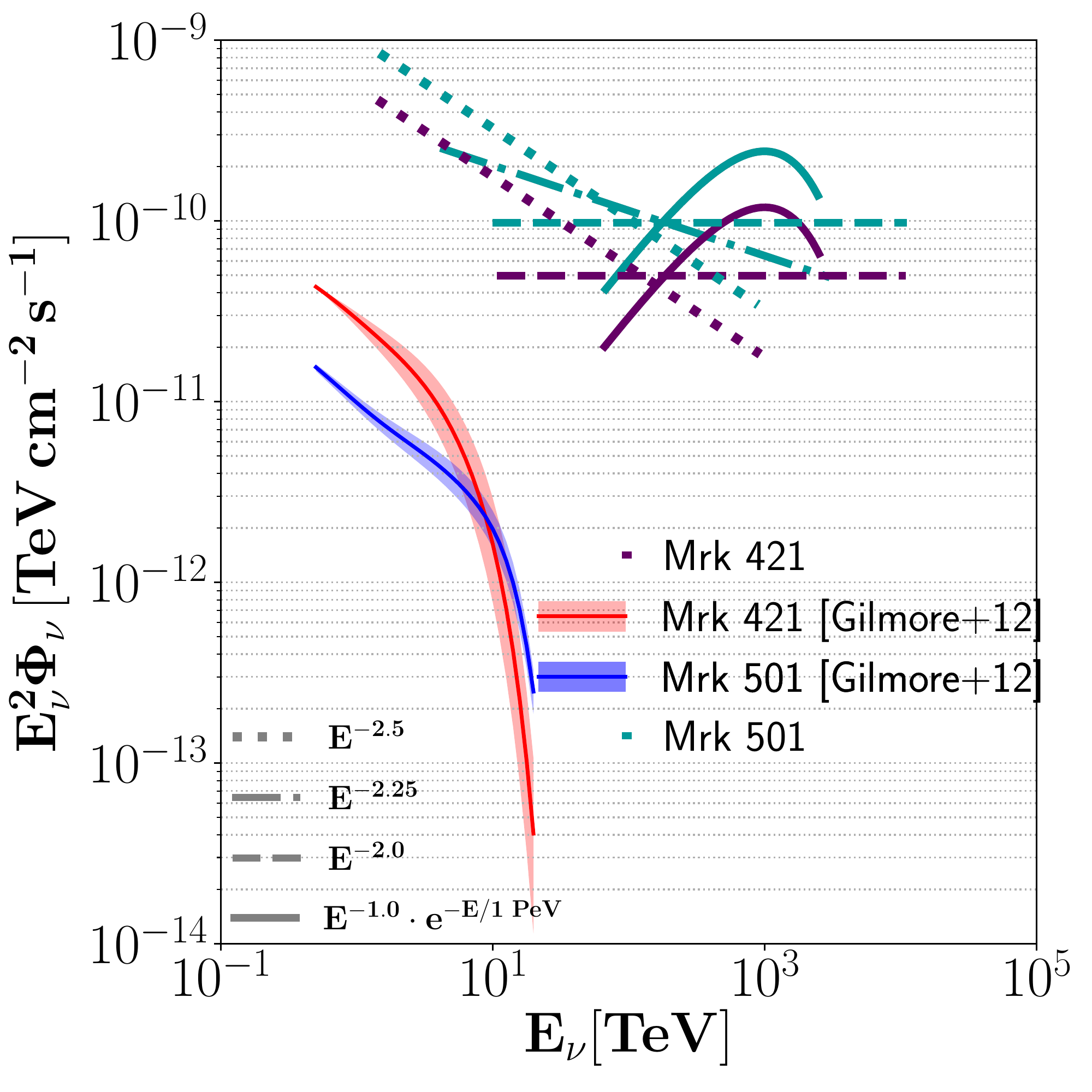} };
        \begin{scope}[x={(image.south east)},y={(image.north west)}]
      
        \node[text width=8cm, minimum height=3cm,minimum width=8cm] at (0.92,0.925) { \textbf{ \textcolor{red}{
        \normalsize{PRELIMINARY}  
       } } };
       \end{scope}
   \end{tikzpicture}
    \end{minipage}%
    \begin{minipage}{0.512\textwidth}
    \vspace{-4.5pt}
        \centering
        \begin{tikzpicture}
        \node[anchor=south west,inner sep=0] (image) at (0,0) 
        {\includegraphics[width=1.0\textwidth]{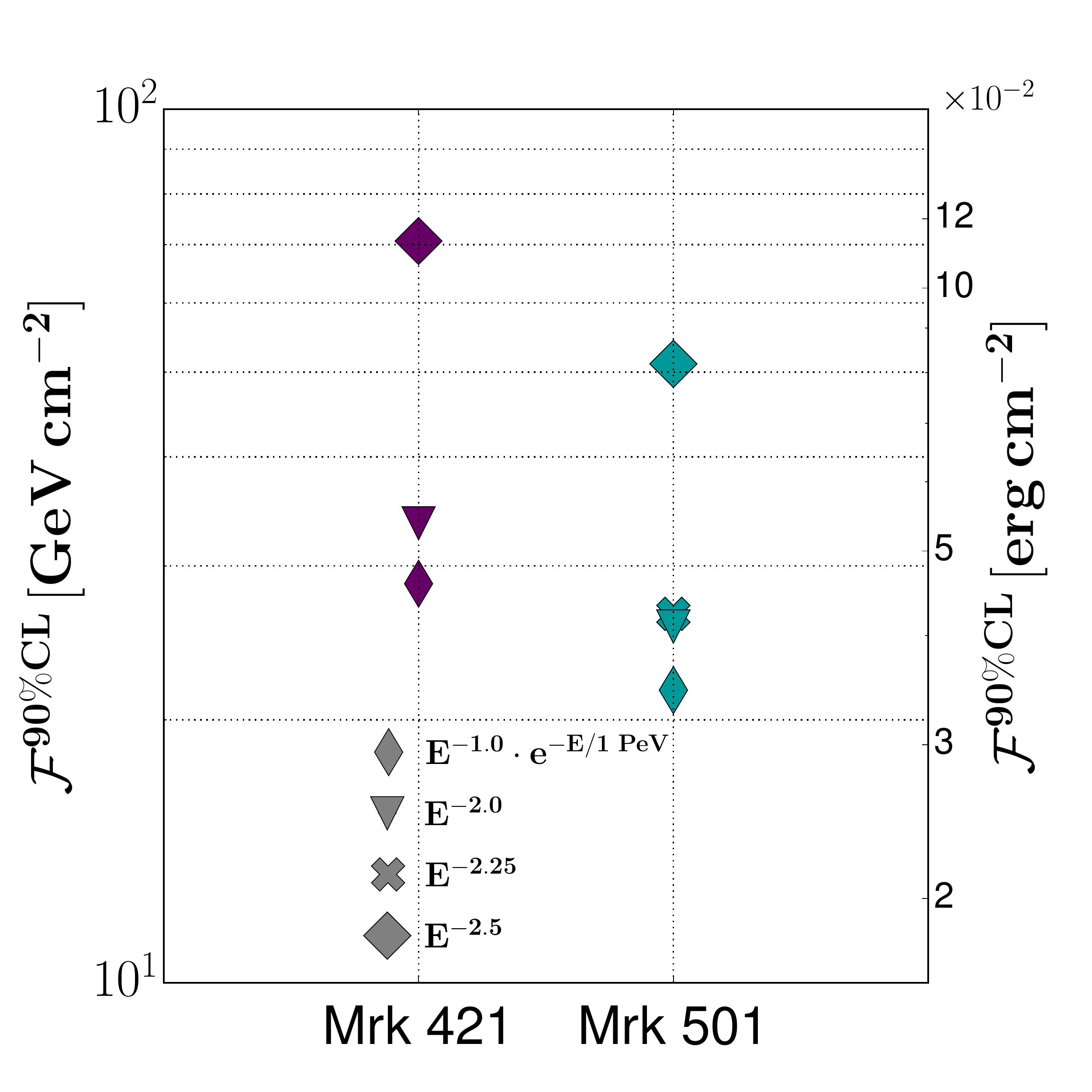} };
        \begin{scope}[x={(image.south east)},y={(image.north west)}]

        \node[text width=8cm, minimum height=3cm,minimum width=8cm] at (0.82,0.86) { \textbf{ \textcolor{red}{
        \normalsize{PRELIMINARY}  
       } } };
       \end{scope}
   \end{tikzpicture}
    \end{minipage}
    \caption{Upper limits on the flux $F$ (left) and fluence $\mathcal{F}$ (right) for different spectra for both blazars. The attenuated fluxes for blazars with the intrinsic spectra obtained in HAWC~\cite{ICRC2019SaraDeCoutino} are shown.}
    \label{fig:ULfluxLandfluenceS2}
\end{figure}

The Figure~\ref{fig:ULfluxLandfluenceS2} gathers the 90\% CL ULs on $\nu$ flux and fluence. The best 90\% CL flux ULs (see Figure~\ref{fig:ULfluxLandfluenceS2}, left plot) is obtained with the case of all flare states selected, while the best 90\% CL fluence ULs (see Figure~\ref{fig:ULfluxLandfluenceS2}, right plot) is obtained with \textit{average flux}+$\mathit{2\sigma}$ threshold.

\begin{figure}[htbp!]
\vspace{-10pt}
\centering
        \begin{tikzpicture}
        \node[anchor=south west,inner sep=0] (image) at (0,0) 
        {\includegraphics[width=1.0\textwidth]{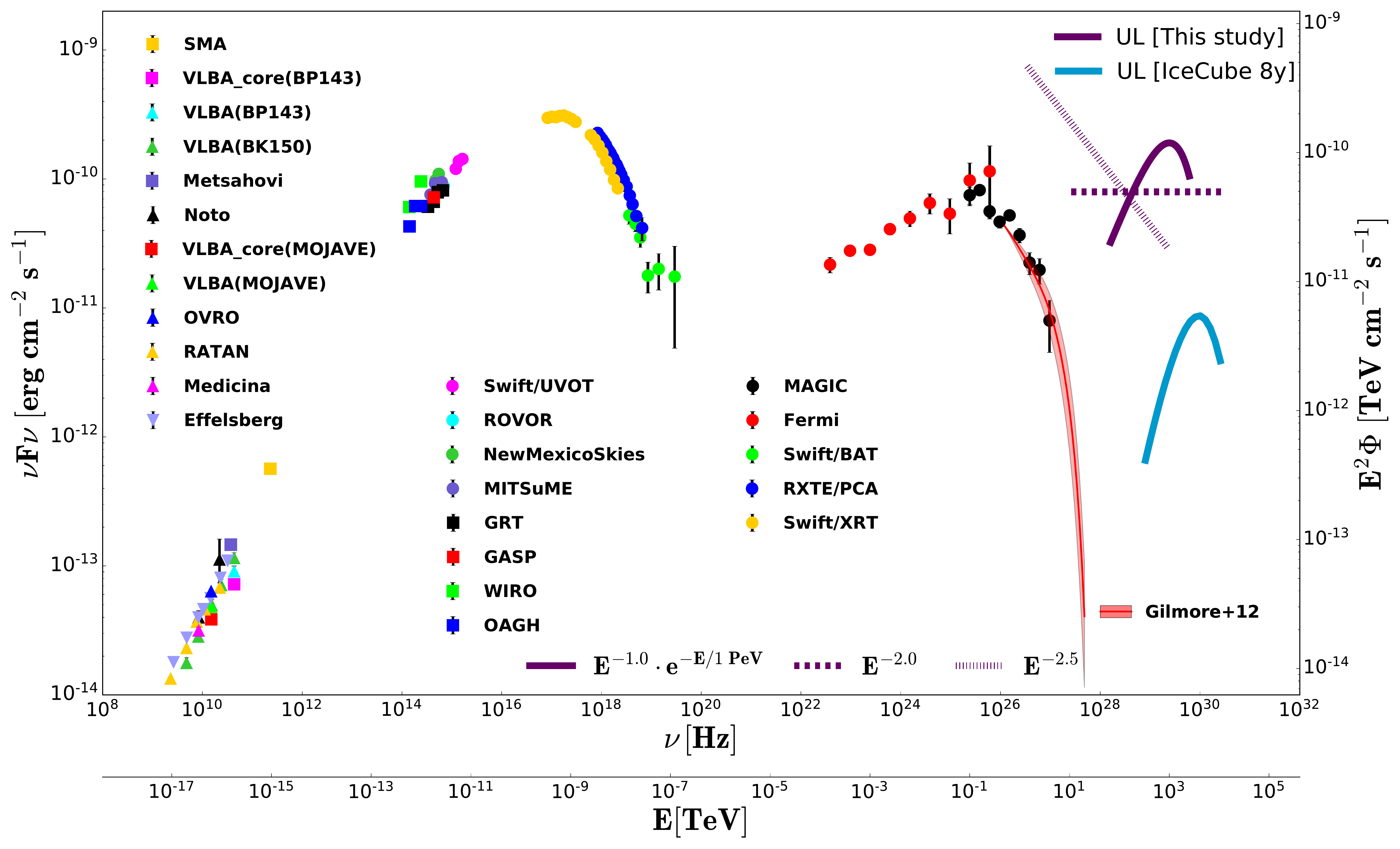} };
        \begin{scope}[x={(image.south east)},y={(image.north west)}]
        \node[text width=8cm, minimum height=3cm,minimum width=8cm] at (0.69,0.95) { \textbf{ \textcolor{red}{
        \normalsize{PRELIMINARY}  
       } } };
       \end{scope}
   \end{tikzpicture}
\caption{Neutrino flux ULs vs $\gamma$-ray SED of Mrk 421. The SED of Mrk 421 averaged over all the observations taken during the multifrequency campaign in 2009. See~\cite{LAT:2011aa} for references to the different instruments involved in the campaign. Adapted from~\cite{LAT:2011aa}, credit by David Paneque. The violet lines indicate the $\nu$ ULs for different spectra obtained in this study. The ULs for IceCube search~\cite{Aartsen:2018ywr} with the default spectrum for non-flaring (quiescence) period obtained from~\cite{Petropoulou:2015upa} is shown (solid blue) for comparison, credit by Ren\'e Reimann. The attenuated flux for Mrk 421 with the intrinsic spectrum obtained in HAWC~\cite{ICRC2019SaraDeCoutino} is shown.}
\label{fig:SEDandMrk421}
\end{figure}

\clearpage 
\begin{figure}[ht!]
\vspace{-25pt}
\centering
        \begin{tikzpicture}
        \node[anchor=south west,inner sep=0] (image) at (0,0) 
        {\includegraphics[width=1.0\textwidth]{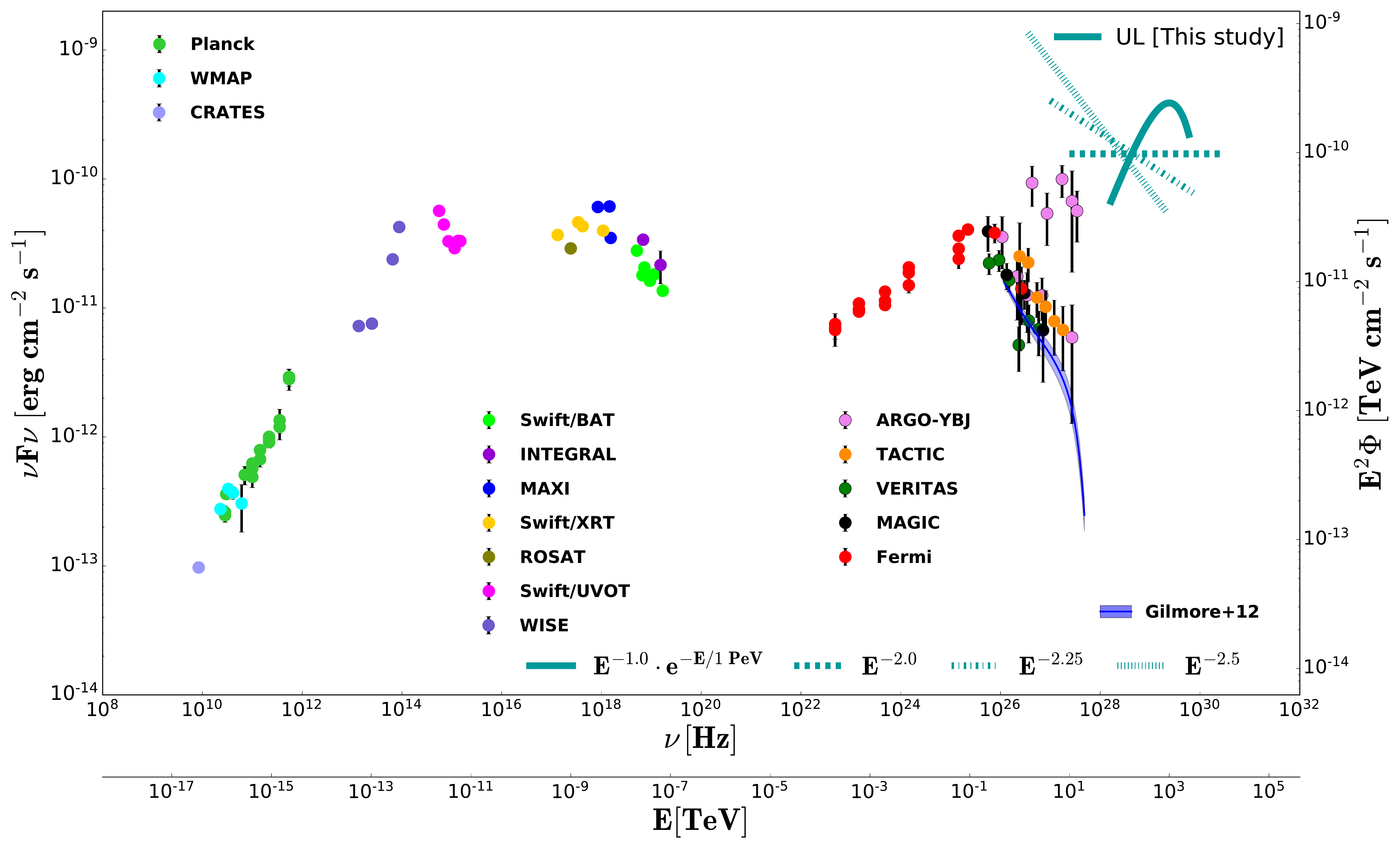} };
        \begin{scope}[x={(image.south east)},y={(image.north west)}]

        \node[text width=8cm, minimum height=3cm,minimum width=8cm] at (0.69,0.95) { \textbf{ \textcolor{red}{
        \normalsize{PRELIMINARY}  
       } } };
       \end{scope}
   \end{tikzpicture}
   \caption[Neutrino flux ULs vs $\gamma$-ray SED of Mrk 501. The data (which based on~\cite{SEDMRK501all}) for multi-frequency SED is gathered from the historical observations available in the SED Builder Tool of the ASI Space Science Data Center (SSDC), which combines radio to $\gamma$-ray band data from several missions and experiments together. The light sea green lines indicate the $\nu$ ULs for different spectra obtained in this study.]{Neutrino flux ULs vs $\gamma$-ray SED of Mrk 501. The data (which based on~\cite{SEDMRK501all}) for multi-frequency SED is gathered from the historical observations available in the SED Builder Tool of the ASI Space Science Data Center (SSDC)\footnotemark, which combines radio to $\gamma$-ray band data from several missions and experiments together. The light sea green lines indicate the $\nu$ ULs for different spectra obtained in this study. The attenuated flux for Mrk 501 with the intrinsic spectrum obtained in HAWC~\cite{ICRC2019SaraDeCoutino} is shown.}
\label{fig:SEDandMrk501}
\end{figure}
\footnotetext{\href{https://tools.ssdc.asi.it/SED/}{https://tools.ssdc.asi.it/SED/}}

\vspace{-10pt}
\section{Conclusions and Outlook}
\label{sec:conc}
The HAWC detector operates nearly continuously and it is currently the most sensitive wide field-of-view $\gamma$-ray telescope in the very promising HE band from 100 GeV to 100 TeV. Therefore, it opens prospects to study the most energetic astrophysical phenomena in the Universe as well as to understand the mechanisms that power them and endeavor to break the mystery of their origin. Taking into account the flare timing information given by $\gamma$-ray observations should improve the efficiency of the search for a $\nu$ counterpart with ANTARES. With the expected decommissioning of ANTARES at the end of 2019, the next-generation multi-km$^{3}$-sized Cubic Kilometer Neutrino Telescope (KM3NeT)\footnote{KM3NeT Collaboration, \href{https://www.km3net.org/}{https://www.km3net.org/}}~\cite{Adrian-Martinez:2016fdl} $\nu$ telescope with a unique design of multi-PMT optical modules will take up with new vigor the challenges faced by ANTARES and will have surpassed it in the sensitivity in few years raising a new era in $\nu$ astronomy. The KM3NeT will provide more than an order of magnitude improvement in sensitivity~\cite{ICRC2019HighlightTalkConiglione}; therefore, such sources are promising candidates as high-energy $\nu$ emitters for an improved future time-dependent search. As shown in~\cite{Petropoulou:2016ujj} for Mrk 421, the muon $\nu$ event rate during a short flaring period is comparable to the one expected from a longer but non-flaring period, i.e. during quiescence; thereby, a collection of flares over a several years is essential to produce a meaningful signal in modern multi-km$^{3}$-sized $\nu$ telescopes.

\end{document}